\documentclass[twocolumn,prb,aps,showpacs]{revtex4}

\usepackage{graphicx}
\def\be{\begin{equation}}
\def\ee{\end{equation}}
\def\bea{\begin{eqnarray}}
\def\eea{\end{eqnarray}}

\begin{document}

\title{Transverse magnetic field distribution in the vortex state of noncentrosymmetric
superconductor with O symmetry}

\author{Chi-Ken~Lu and Sungkit~Yip}

\affiliation{Institute of Physics, Academia Sinica, Nankang,
Taipei 115, Taiwan}

\date{\today }

\begin{abstract}
We investigate the magnetic field distribution inside a Type II
superconductor which has point group symmetry $O$ such as
Li$_2$Pt$_3$B. The absence of inversion symmetry as a departure from
perfect cubic group $O_h$ causes a magnetization collinear with the
phase gradient associated with the order parameter, and a component
of current collinear with the local magnetic field. In the vortex
state, we predict, by solving the Maxwell equation, a local magnetic
field transverse to the vortex lines. The probability distribution
of this transverse field is also obtained for the prospective
muon-spin-rotation measurements.


\end{abstract}


\maketitle

\section{Introduction}

Noncentrosymmetric heavy-fermion superconductors have recently
triggered much theoretical and experimental attention. The family of
Li$_2$B(Pd$_{1-x}$Pt$_x$)$_3$ as well as the compounds CePt$_3$Si
and CeRhSi$_3$ are among the major examples.
\cite{Togano,exp_Li,Nishiyama,Badica,Bauer,Yasuda,Kaur,Ce,Onuki}
The lack of inversion symmetry of background crystal causes an
unconventional pairing in the superconducting state and a number of
consequences are predicted.\cite{Gorkov,review_th1,FFLO,Samokhin} It
is commonly believed that in the normal state the Fermi surface is
split by the spin-orbital coupling as a result of the broken
inversion symmetry, and the pairing wavefunction is in a mixture of
singlet and triplet pseudospin state.\cite{Gorkov} Suggested by the
low temperature measurements of penetration depths, the family of
Li$_2$B(Pd$_{1-x}$Pt$_x$)$_3$ with point group symmetry $O$
demonstrates a trend from a conventional gap to a nodal gap caused
by mixing with a stronger triplet pairing as the composition $x$
varies from 0 to 1.\cite{exp_Li} The conclusion that the
Li$_2$Pd$_3$B compound is a fully gapped superconductor is
consistent with other measurements such as NMR,\cite{NMR} specific
heat,\cite{spheat} and muon spin rotation ($\mu$SR)
measurements.\cite{muSR,Hafliger} The nodal gap structure of
Li$_2$Pt$_3$B hinted by the penetration depth measurement, however,
is controversial since the specific heat and $\mu$SR measurements
suggest conventional pairing.\cite{spheat,Hafliger}


Microscopically, the lack of inversion symmetry allows a
parity-breaking spin-orbital interaction which leads to a
magnetization induced by the phase gradient associated with the
superconducting order parameter, and vice versa.\cite{Edelstein} For
an ideally two-dimensional system, the phase gradient induced by the
constant magnetic field, or the so-called helical phase, might be
measured using a Josephson junction between a conventional and a
noncentrosymmetric superconductor.\cite{Kaur} However, this scheme
is not feasible when considering the decay of magnetic field in a
real three-dimensional bulk superconductor. On the other hand, a
unique and macroscopic distribution of magnetic field is generated
inside the noncentrosymmetric superconductors due to the induced
phase graident and magnetization.\cite{Levitov,Oka,Yip,Lu}
Therefore, the measurement of the local field distribution, such as
the $\mu$SR,\cite{SLee} can be used to detect the unconventional
pairing. Given the fact that these noncentrosymmetric compounds are
strong Type II superconductors,\cite{Ce,Badica} the effects of
lacking inversion symmetry on the vortex state are of importance. In
this paper we focus on the noncentrosymmetric superconductors with
point group symmetry $O$, such as Li$_2$Pt$_3$B, where a
distribution of transverse magnetic field is generated in addition
to the ordinary field parallel with the vortex line.\cite{Lu}


In Sec.\ref{singleVTEX}, we first derive the magnetic distribution
associated with an isolated vortex line. In particular, we note a
singular bound current defined by the curl of magnetization
appears within the vortex core. It appears because the
magnetization here is proportional to the gradient of order
parameter phase associated with the vortex state. Hence a new
boundary condition is necessary for the azimuthal field to balance
the singular bound current, which was missing in our previous
study,\cite{Lu} and it leads to a significant correction to this
field. In Sec.\ref{lattice}, we extend to the lattice and
numerically obtain the tranverse field $B_{\perp}(x,y)$ as well as
the associated probability distribution $p(B_{\perp})$. In
Sec.\ref{conclusion} we connect the present results with the
proposed helical phase, and draw the conclusion in the end.

\section{One isolated vortex line}\label{singleVTEX}

Inside a type II conventional superconductor in the presence of
external magnetic field, the Abrikosov vortex state with a periodic
distribution of internal magnetic field is developed when the
applied field is larger than $H_{c1}$, below which a complete
Meissner effect occurs, and less than $H_{c2}$, beyond which the
superconductivity only appears on the surface and completely
disappear for further increased field. The internal magnetic field
inside a conventional superconductor is parallel with the external
one, whereas a transverse component is developed inside this
noncentrosymmetric superconductor, as will be shown below. Since the
internal field for an isolated vortex is essential to the vortex
lattice, below we first solve the single vortex case.

Consider a noncentrosymmetric superconductor with an isolated vortex
line along the z direction. Below the cylindrical coordinate
$(\rho,\phi,z)$ is employed. We denote the order parameter
associated with the single vortex as $|\psi|e^{i\chi}$ where the
phase $\chi$=$-\phi$, that is the vortex is said to have the winding
number of minus one. Besides, the magnitude $|\psi|$ can be viewed
as a constant for positions outside the vortex core that roughly has
a size of coherence length $\xi$. The internal local magnetic field
$\bf{B}$ is then determined by the supercurrent $\bf{J}$ through the
Amp\`{e}re's law,

\be
    \nabla\times{\bf{B}}=4\pi\nabla\times{\bf{M}}
    +\frac{4\pi}{c}(-e)\bf{J}\:,\label{Max1}
\ee where the first term on the right represents the bound current
due to the magnetization $\bf{M}$, which is zero for conventional
superconductors. The electric charge has been written as $(-e)$.
Here
the nonzero $\bf{M}$ is proportional to the phase
gradient,\cite{Edelstein,Levitov} namely,

\be
    4\pi{\bf{M}}=\frac{\delta}{{\lambda}^2}
    ({\bf{A}}+\frac{c}{2e}\nabla\chi)\label{magnetization}\:,
\ee where $\lambda$ is the London penetration depth, $\delta$ has
the dimension of length, and the small number $\delta/\lambda \equiv
\tilde{\kappa}/2$ depends on the strength of the spin-orbital
interaction associated with the lack of inversion symmetry.\cite{Lu}
On the other hand, the internal magnetic field also induces a
contribution to the supercurrent, which together with the ordinary
contribution can be written as,

\be
    \frac{4\pi}{c}e{\bf{J}}=\frac{1}{{\lambda}^2}
    ({\bf{A}}+\frac{c}{2e}\nabla\chi)
    -\frac{\delta}{{\lambda}^2}{\bf{B}}\:,\label{current}
\ee where $\bf{A}$ is the vector potential defining the internal
field. Note that $\hbar$=1 for convenience. ${\bf{M}}$ in eq
(\ref{magnetization}) and the last term in eq.\ (\ref{current})
exist only if the superconductor is non-centrosymmetric. We note
here that \cite{Lu} both $1/\lambda^2$ and $\delta / \lambda^2$ are
proportional to $1 - Y(T)$, where $Y(T)$ is the Yosida function, so
$\tilde \kappa$ is $\propto \left( 1 - \frac{T}{T_c} \right)^{1/2}$
near $T_c$ and increases with decreasing temperature, saturating to
a constant value as $T \to 0$.

The solutions to Eq.\ (\ref{Max1}) go back to the standard singly
quantized vortex when $\tilde{\kappa}$ is set to be zero. The
corresponding internal field $B_z= B_z^{(0)} \equiv
\frac{\Phi_0}{2\pi\lambda^2}K_0(\rho/\lambda)$ where $\Phi_0$ is a
quantum of flux $\frac{\pi\hbar{c}}{e}$ and $K_0$ is the modified
Bessel function of zeroth order. For nonzero $\tilde{\kappa}$, an
azimuthal internal field $B_{\phi}$ emerges. In addition, near the
core the magnetization contains an azimuthal component proportional
to $1/\rho$ from the phase gradient term in eq.\
(\ref{magnetization}). This diverging behavior of $M_{\phi}$ results
in a diverging $B_{\phi} \propto 1/\rho$, which was unfortunately
left out in our earlier work.\cite{Lu} To see this in more detail,
we first note that, due to the Meissner effect, we expect that ${\bf
B}$, the gauge invariant velocity $(\nabla \chi + \frac{2e}{c} {\bf
A})$ hence also ${\bf M}$, ${\bf J}$ all decay to zero as $\rho \to
\infty$.  We note that this guarantees the total magnetic flux
associated with a singly quantized vortex remains to be $\Phi_0$, as
$\oint ({\bf A} + \frac{c}{2e} \nabla \chi) \cdot d {\vec l} = 0$
over a large loop encircling the vortex. Integrating eq (\ref{Max1})
over a large area enclosing the vortex and using Stokes theorem, we
obtain

\be \oint {\bf B} \cdot d {\vec l} - 4 \pi \oint {\bf M} \cdot d
{\vec l} = \frac{4 \pi} {c} (-e) \oint {\bf J} \cdot {\bf dS}\:,
\label{jz} \ee

\noindent hence automatically the condition of zero total transport
current along the vortex line, since $B_{\phi}$, $M_{\phi}$ decays
to zero exponentially at large $\rho$.

However, considering eq (\ref{jz}) for a small loop encircling the
vortex, we see that for $\rho \to 0$, $B_{\phi} = 4 \pi M_{\phi}$.
Using eq (\ref{magnetization}) and the fact that the magnetic flux
passing through a small loop must necessarily approaches zero when
the size of the loop does, we get $ 4 \pi M_{\phi} \times 2 \pi \rho
= -\frac{\delta}{\lambda^2} \Phi_0$ hence

\be
    B_{\phi}(\rho\rightarrow\xi)=-\frac{\tilde{\kappa}}{2\lambda}
    \frac{\Phi_0}{2\pi\rho}\: ,\label{newBC}
\ee

This behavior can also be obtained directly by substituting eq
(\ref{magnetization}) and (\ref{current}) into (\ref{Max1}) and
using $\nabla\times\nabla\chi = - 2 \pi
\delta^{(2)}(\vec{\rho})\hat{z}$.  We get

\be
    \nabla\times{\bf{B}}=\frac{4\pi}{c}(-e){\bf{J}}
    +\frac{\tilde{\kappa}}{2\lambda}
    \left[{\bf{B}}-\Phi\delta^{(2)}(\vec{\rho})\hat{z}\right]
    \: .\label{Max2}
\ee  The delta function  can be taken care of by an appropriate
boundary condition associated with the azimuthal field $B_{\phi}$ at
the origin, as in eq (\ref{newBC}).

Eq.\ (\ref{Max2}) can be solved in the manner of perturbation in
$\tilde{\kappa}$. For $\rho\neq0$, the delta function can be ignored
and the first-order correction $B^{(1)}_{\phi}$ satisfies the
following equation, obtained by taking the curl of eq.\ (\ref{Max2})
and using eq.\ (\ref{current}),

\be
    \left[\frac{d}{d\rho}(\frac{1}{\rho}\frac{d}{d\rho}\rho)
    -\frac{1}{\lambda^2}\right]B^{(1)}_{\phi}(\rho)=
    \frac{\tilde{\kappa}}{\lambda}
    \frac{d}{d\rho}B^{(0)}_z\:,\label{vortex2}
\ee in which the source term on the RHS arises from the extra
supercurrent induced by the unperturbed internal field $B^{(0)}_z$
as well as the second term in eq. (\ref{Max2}). This equation is
identical to eq. (37) in our previous work. \cite{Lu} There the
inhomogeneous solution satisfying $B_{\phi}\rightarrow0$ near the
origin was obtained and we shall denote it as
$\frac{\tilde{\kappa}\Phi_0}{2\pi\lambda^2}f(x)$ with the argument
$x$ defined by $x=\rho/\lambda$. (Thus $f(x)$ represents the right
hand side of eq (41) in Ref.\ 22). We note that this inhomogeneous
solution does not satisfy the boundary condition in eq.\
(\ref{newBC}), which can be fixed by adding an appropriate
homogeneous solution to eq.\ (\ref{vortex2}) satisfying eq
(\ref{newBC}).  Since $K_1(x)$ fulfills eq.\ (\ref{vortex2}) in the
absence of the source term and $K_1 (x) \propto 1/x$ for small $x$,
we can easily write down the desired azimuthal field as,

\be
    B^{(1)}_{\phi}(x)=\frac{\tilde{\kappa}\Phi}{2\pi\lambda^2}
    \left[
    f(x)-\frac{1}{2}K_1(x)
    \right]\:.\label{Solution}
\ee The azimuthal field still decays exponentially at infinity,
justifying the previous assumptions. However, in contrast to our
previous results,\cite{Lu} due to the $K_1$ term the azimuthal field
for an isolated vortex line now has increasing magnitude towards the
core.

\section{lattice of vortex lines}\label{lattice}

\begin{figure}
\input{epsf}
\epsfxsize=3in \epsfysize=3in \epsfbox{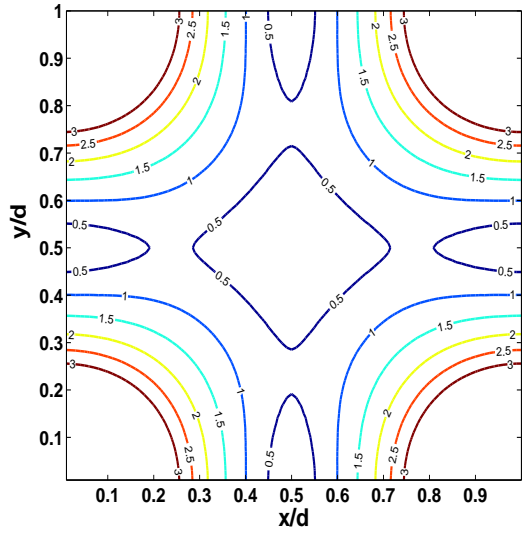}
\caption{Contour plot of transverse magnetic field magnitude
$B_{\perp}$ for a square lattice. The unit of $B_{\perp}$ is
$B_0=\tilde{\kappa}\Phi/2\pi\lambda^2$. The distance $d$ between
neighboring lines to the penetration depth $\lambda$ is
0.5.}\label{Square}
\end{figure}

\begin{figure}
\input{epsf}
\epsfxsize=3.5in \epsfysize=2.1in \epsfbox{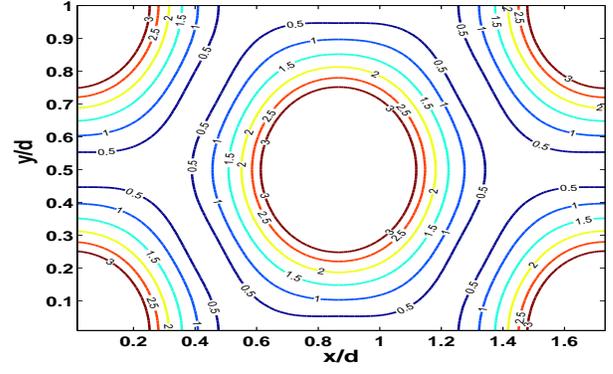}
\caption{The same contour plot as Fig. \ref{Square} but for a
triangular lattice with the same value of
$d/\lambda$.}\label{Triangular}
\end{figure}

Since those heavy-fermion superconductors usually have low
transition temperatures and large Ginzburg-Landau parameter,
$H_{c1}$ is practically small and usually much less than the
external magnetic field, which causes the superconductor to be in
the vortex lattice state in most cases. The components for
transverse magnetic fields $B_x(x,y)$ and $B_y(x,y)$ inside the
superconductor are given by the vector sums of eq.\ (\ref{Solution})
over the vortex lines located at each of the lattice vertex. For the
family of Li$_2$B(Pd$_{1-x}$Pt$_x$)$_3$, the ratio of $\lambda/\xi$
is about\cite{Badica} 20 independent of the Pt composition. The
transverse field depends on the lattice geometry, and $d/\lambda$ is
the only parameter with $d$ being the lattice constant. Figs.\
\ref{Square} and \ref{Triangular} show the contours of transverse
field $B_{\perp}=\sqrt{B_x^2+B_y^2}$ in unit of
$B_0=\tilde{\kappa}\Phi_0/2\pi\lambda^2$  with $d/\lambda$=0.5 for a
square and triangular lattices, respectively. Due to the delta
function in eq.\ (\ref{Max2}), the regions close to the core have
larger transverse fields. The destructive interferences from
neighboring vortex lines occur at the middle points on each side and
at the center of each square and triangle, respectively.

\begin{figure}
\input{epsf}
\epsfxsize=3in \epsfysize=3in \epsfbox{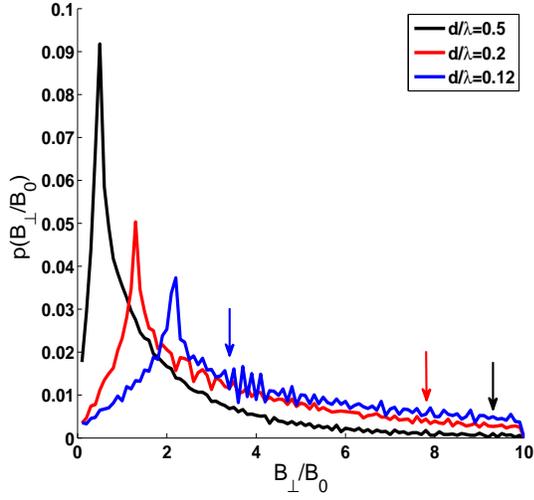} \caption{The
probability distribution $p(B_{\perp}/B_0)$ of internal transverse
fields for square lattice with various values of $d/\lambda$. The
arrows indicate the cutoff magnetic fields associated with the
vortex core. The respective cutoff values are 9.6, 7.9, and 3.5
B$_0$.}\label{SquarepB}
\end{figure}

\begin{figure}
\input{epsf}
\epsfxsize=3in \epsfysize=3in \epsfbox{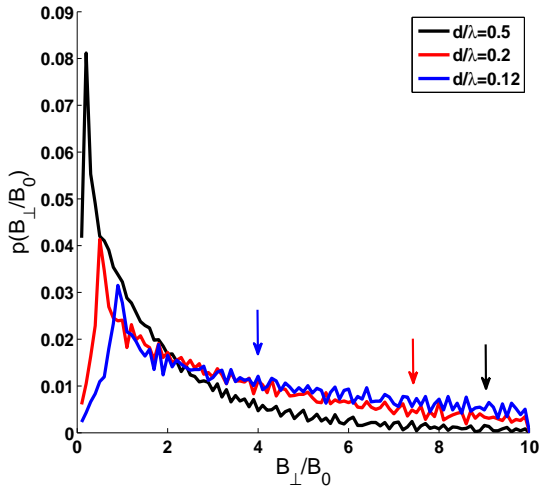}
\caption{$p(B_{\perp}/B_0)$ for triangular lattice for various
values of $d/\lambda$. The respective cutoff values are 9.2, 7.7,
and 4.0 B$_0$.}\label{TripB}
\end{figure}

Statistics on the local internal field $B_{\perp}(x,y)$ give the
probability distribution $p(B_{\perp}/B_0)$. In Fig.\
\ref{SquarepB} and Fig.\ \ref{TripB}, we obtain the probability
distributions $p(B_{\perp}/B_0)$ for various values of $d/\lambda$
in the square and triangular lattices, respectively.
$p(B_{\perp}/B_0)$ vanishes for $B_{\perp} \to 0$, and rises up
linearly with $B_{\perp}$ for small $B_{\perp}$.  It reaches a
peak corresponding to the longest contour in Fig. \ref{Square} or
\ref{Triangular}, then decreases with increasing $B_{\perp}$. As
the value of $d/\lambda$ decreases, the vortex lines are getting
denser and the magnetic fields associated with the peak of
$p(B_{\perp})$ become greater. One should note, however, that
there is a cutoff magnetic field associated with the vortex core
for each value of $d/\lambda$ because the minimum length scale for
the above theory to be applicable here is the coherence length
$\xi$. The colored arrows are used to label these cutoffs for
various $d/\lambda$. That is, the values of $B_{\perp}$ beyond
these arrows are un-physical and the actual values are expected to
be much smaller. We note that the cutoff for a single vortex is
approximately given by $\frac{1}{2}K_1(\xi/\lambda)\approx10$
times $B_0$. For low density of vortices, i.e. $d/\lambda\gg$1,
the cutoff remains the same as that of a single vortex. For higher
density cases, these cutoffs are modified due to the neighboring
vortices.

The distributions $p(B_{\perp})$ can in principle be obtained from
$\mu$SR measurements, operated in the following manner. Positive
muons with spin polarizing along the external field and hence the
vortex lines are injected into the superconductor in the vortex
state.  In such a set-up for conventional superconductors, the muon
spin polarization is not affected by the local magnetic fields, as
they are parallel (assuming that the vortex lines are straight, {\it
c.f.} below). The decay product, the positrons, has an anisotropic
distribution, being preferentially along the initial muon spin.
However, for our non-centrosymmetric superconductor, the muon spins
will precess around the local field $B_{\perp}(x,y)$. This gives
rise to an extra decay or oscillation of the anisotropic positron
distribution.  This distribution is further a function of external
field and temperature, with magnetic field magnitude increasing with
decreasing temperature if below $T_c$. These features may be used to
distinguish from other contributions. Thermal fluctuation of vortex
lines may also give rise to horizontal fields, but they would
increase instead of decrease with temperature. Magnetic fields from
nuclear moments would have a temperature dependence uncorrelated
with $T_c$.

The values of $d/\lambda$ correspond to average external magnetic
fields $B_{av}$ according to the relation
$B_{av}$=$\frac{2\pi}{\eta}\frac{\lambda^2}{d^2}\frac{\Phi_0}{2\pi\lambda^2}$,
where $\eta$ is a factor depending on the lattice geometry. For
square lattice $\eta$=1, and for triangular lattice
$\eta$=$\frac{\sqrt{3}}{2}$. The magnetic field
$\frac{\Phi_0}{2\pi\lambda^2}$ is about $25$ Gauss given $\lambda(T
\to 0)$=360nm in Li$_2$Pt$_3$B.\cite{Badica} If the parameter
$\tilde{\kappa}$ exceeds 10$^{-2}$, then the typical fields
according to Fig \ref{SquarepB} and \ref{TripB} will be of order 1
gauss, which is supposed to be measurable in the $\mu$SR
measurement.\cite{RK,Schenck} In the compound Li$_2$Pt$_3$B,
however, $\tilde{\kappa}$  may unfortunately be of order 10$^{-3}$
only, with transverse fields that are hard to detect. Nevertheless,
our calculations apply to any noncentrosymmetric superconductor with
O symmetry. The transverse fields may be easily measurable if
$\tilde \kappa$ is a few times larger and/or $\lambda$ a few times
shorter.

\section{discussion}\label{conclusion}

The above paragraphs demonstrate the transverse magnetic field in
the vortex state as a signature of broken inversion symmetry. Here
we elaborate on the general phase gradient
$\vec{Q}=\nabla\chi+\frac{2e}{c}{\bf{A}}$ since this vector is
associated with the helical phase which is described by an order
parameter $|\Psi|e^{i{\vec{Q}}\cdot\vec{r}}$. In a strictly
two-dimensional system with open boundary, the phase gradient can be
obtained by directly setting the current density in eq.\
(\ref{current}) to be zero. Since there is no Meissner effect in
such an ideal system, $\vec{Q}$ is determined solely by the external
field and was argued to be measurable by a Josephson junction
between a conventional and a noncentrosymmetric
superconductor.\cite{Kaur}

For our three-dimensional superconductor, $\vec{Q}$ can be obtained
from eq.\ (\ref{Max1}) and (\ref{current}) given the internal
magnetic field ${\bf{B}}$.  For positions not exactly at the vortex
cores,

\be
    \frac{c}{2e}\vec{Q}=
    \lambda\tilde{\kappa}{\bf{B}}-\lambda^2(\nabla\times{\bf{B}})\:.
\ee In a real bulk superconductor, it can be seen that the phase
gradient is small and not appreciable due to the Meissner effect.
For example, consider a noncentrosymmetric superconductor occupied
by a single vortex. The unique phase gradient $Q_z$ is along the
line. Given the internal fields $B_z^{(0)}$ and $B^{(1)}_{\phi}$
obtained previously, $Q_z$ is vanishingly small at positions $\rho
\gg \lambda$.  Only near the vortex $Q_z$ is large and given by
$\frac{\tilde{\kappa}\ln{x}}{2\lambda}$ for $\rho \ll \lambda$. We
note however that, at these distances, $Q_{\phi} \to -
\frac{1}{\rho}$, hence $|Q_{\phi}| \gg |Q_{z}|$. The phase gradient
associated with the helical phase in this noncentrosymmetric
superconductors has also been obtained beyond the present modified
London theory \cite{Samokhin} (see also \cite{Kaur}).

\section{conclusion}

We consider the effects of broken inversion symmetry
noncentrosymmetric superconductor with O symmetry such as
Li$_2$Pt$_3$B by calculating the transverse magnetic field in the
vortex state. The probability distribution of the field is also
calculated for the prospective $\mu$SR experiments as a signature of
the broken inversion symmetry.

\begin{center}
{\bf Acknowledgment}
\end{center}

We thank Rustem Khasanov and Baruch Rosenstein useful
correspondences. This work is supported by the National Science
Council of Taiwan, R.O.C. under grant No.NSC95-2112-M001-054-MY3.

\newpage

\end{document}